# Transparency and stability of low density stellar plasma related to Boltzmann statistics, inverse stimulated bremsstrahlung and to dark matter

Y. Ben-Aryeh

Technion-Israel Institute of Technology, Physic Department, Israel, Haifa, 32000

Email:   phr65yb@technion.physics.ac.il

**ABSTRACT**

The rate of stimulated inverse bremsstrahlung is calculated for low electron density stellar plasmas and the condition under which the plasma becomes transparent is presented. The stability of low density stellar plasma is analyzed for a star with a spherical symmetry in equilibrium between the gravitational attractive forces and the repulsive pressure forces of an ideal electron gas where the analysis is developed by the use of Boltzmann statistics. Fundamental and surprising results are obtained by which the radius and the total mass of the star are inversely proportional to the square root of the electron density in the star center. The total gravitational forces of the star with very low electron and mass densities are very large (!) due to the extreme large star volumes. The absorption and emission of radiation for extremely low density star plasmas vanishes over all the entire electromagnetic spectrum. The present results are supported by numerical calculations. Similar effects are predicted for low density stellar plasmas which have different structures and the properties of such plasmas might show certain similarities with those of dark matter.

**Key words.**   Stellar plasma with low electron densities-inverse stimulated bremsstrahlung-dark matter

## 1. Introduction

The stellar plasmas statistics under non-relativistic conditions can be analyzed for two different asymptotic regions: a) The Rutherford-Born approximation is valid for very fast electrons giving classical collision frequency which scales as $1/v^3$ with electron velocity $v$ [e.g. Ginzburg 1961; Piel 2010; Chen 1990; Fitzpatrick 2015]. The use of Born approximation is valid under the conditions: $Ze^2/4\pi\varepsilon_0\hbar v \ll 1$, $m_e v^2/2\hbar\omega \gg 1$, where $e$ is the electron charge, $\hbar$ the Planck constant (divided by $2\pi$), $m_e$ the electron mass and $\omega$ the frequency of the electromagnetic field.( e.g. Marcuse 1962; Krainov 2000; Meyer-ter–Vehn 2008). (In the present work we use the MKS units and treat the plasma with one ionic component but the analysis can easily be generalized to more ionic components). Very dense stellar plasmas satisfy Fermi-Dirac statistics where the average electron velocity is depending only on the electrons density and is independent of temperature [e.g.  Ashcroft & Mermin 1967;  Seeger 1991].  For this case by averaging the dependence on $1/v^3$ over Fermi-Dirac distribution we obtain strong absorption and strong reflection for light incident on such plasmas. Such conditions are valid for fast electrons with high densities. b) In the present work we are interested in the case of relatively slower electrons and smaller densities for which    $Ze^2/4\pi\varepsilon_0\hbar v > 1$. The absorption in this case is equal to bremsstrahlung stimulated absorption minus the stimulated bremsstrahlung emission. In zeroth order approximation the total rate of stimulated radiation for one electron vanishes [Krainov 2000]. We calculate the rate of stimulated inverse absorption in first order approximation by taking into account the difference in the number of electrons for stimulated emission with those for stimulated absorption and find the conditions under which the plasma becomes nearly transparent. Very high densities of plasmas producing white dwarfs, which are related to Fermi-Dirac statistics, were treated quite long ago [Chandrasekhar 1967; Edington 1959; Kippenhann, Weigert &Weiss 2013]. In the present work we treat the transparency of plasmas with relatively low densities at enough high temperatures at which Boltzmann statistics is valid. While the mass density of a star with the present plasma conditions is relatively low its stability conditions, derived in the present work, show that it can extend on extremely large volumes so that its gravitational forces become very significant. Due to the transparency of such stars we raise the idea that such star plasmas might have implication to dark matter.

Dark matter is thought to be non-baryonic, possibly being composed of some as-yet undiscovered particles [e.g. Bertone and Hooper 2018; Trimble 1987]. Its presence is implied in astrophysical observations, due to the existence of gravitational effects that cannot be explained unless more matter is present that cannot been seen. The name dark matter refers to the fact that it does not appear to interact with electromagnetic radiation and thus it is invisible ('dark') to the electromagnetic spectrum. Usually stars are observed either by radiation emitted by them or by reflected and transmittance of light incident on them from other sources. It is interesting to note by following the present analysis that stellar plasmas with very low densities do not emit any significant radiation and their transmittance is nearly equal 1 (reflectance nearly zero). We find that the stability of a star under spherical symmetry, which is based on the present plasmas conditions, can extend on extremely large volumes so that its gravitational forces become very large. While we developed the present analysis for stars with spherical symmetry similar results are predicted for star plasmas which have different structures. We study the possibility that such plasmas might be related to dark matter.

The present paper is arranged as follows:

By using the theories on stimulated inverse bremsstrahlung, we treat in section 2 the transmittance of monochromatic electromagnetic field through stellar plasmas with relatively slow electrons and low densities. We show that under certain conditions analyzed in the present work these plasmas have extremely low transmittance and reflectivity. In section 3 we study the stability conditions of such plasmas showing that that although their mass density is quite low they can extend on extremely large volumes and therefore they produce very significant gravitational forces. Due to these properties we find that such plasmas might be related to dark matter. In section 4 we demonstrate our results by numerical calculations. In the Appendix we develop alternative method for calculating the star radius which gives similar result to that developed in the main article. In Section 5, we summarize our results.

## 2. Optical properties of low densities stellar plasmas related to inverse stimulated bremsstrahlung and to plasma transparency

The differential cross section for stimulated emission or absorption for one electron-ion collision (neglecting quantum corrections) was given by [Krainov 2000; Berestetskii, Lifshitz & Pitaevsky 1998 ; Manakov 2015]:

$$d\sigma = \frac{16\pi}{3\sqrt{3}} \frac{e^2}{4\pi\varepsilon_0 \hbar c} \left( \frac{ze^2}{m_e v^2 4\pi\varepsilon_0} \right)^2 \frac{v^2}{c^2} \frac{d\omega}{\omega} \quad . \tag{1}$$

The term $\frac{e^2}{4\pi\varepsilon_0 \hbar c} \simeq \frac{1}{137}$ in Eq. (1) represents the fine structure constant. Eq. (1) is a good approximation under the condition $\left(\frac{ze^2}{4\pi\varepsilon\hbar v}\right) > 1$ for slow electrons. In Eq. (1) we apply the transformation [Krainov 2000; Fedorov 1997]:

$$\frac{\varepsilon_0 E^2}{2} = \hbar\omega \frac{\omega^2 d\omega}{\pi^2 c^3} \rightarrow d\omega = \frac{\pi^2 c^3 \varepsilon_0 E^2}{2\hbar\omega^3} \qquad . \tag{2}$$

Here $\frac{\varepsilon_0 E^2}{2}$ is the energy density of the electric field $E$, $\frac{\omega^2 d\omega}{\pi^2 c^3}$ is the number of photons per unit volume for the nearly monochromatic electromagnetic field with frequency interval $d\omega$ and $\hbar\omega$ is the energy of one photon [Krainov 2000; Fedorov 1997]. By substituting Eq. (2) into Eq. (1) we get

$$d\sigma = \frac{1}{8\cdot 3\sqrt{3}} \frac{e^6 Z^2}{\varepsilon_0^2 \omega^4 \hbar^2} \frac{E^2}{p^2} \qquad , \tag{3}$$

where $p$ is the electron momentum. The rate of the *stimulated emission,* or *absorption* $w_{e,a}$, of the photon in the Coulomb scattering of a relatively slow electron is given by multiplying $d\sigma$ by $n_i p$ where $n_i$ is the number of ions per unit volume [Krainov 2000]:

$$w_{e,a} = \frac{1}{8\cdot 3\sqrt{3}} \frac{n_i e^6 Z^2}{\varepsilon_0^2 \omega^4 \hbar^2} \frac{E^2}{p} = C\frac{1}{p} \quad ; \quad C = 4.67\cdot 10^{-25} \frac{n_i Z^2 E^2}{\omega^4} \qquad . \tag{4}$$

We notice that $w_{e,a}$ is proportional to $1/p$ in comparison to the dependence on $1/p^3$ for fast electrons. We notice also that $w_e$ for stimulated emission is equal to stimulated absorption $w_a$ so that the total stimulated radiation $w_T$ for one electron-ion collision in zeroth order approximation vanishes, i.e. [Krainov 2000]:

$$w_T = w_e - w_a = 0 ! \qquad . \tag{5}$$

In order to get the total rate of inverse stimulated absorption of the plasma we need to multiply $w_{e,a}$ of Eq. (4) by the number of ionized electrons of stimulated emission from the upper level minus the number of ionized electrons of stimulated absorption from the lower level and average the term $1/p$ over the Boltzmann distribution. The difference between the electron energy $E_{up}$ in the upper level for any stimulated emission and the corresponding energy of the electron in the lower level $E_{down}$ for the inverse transition of stimulated absorption is given by $E_{up} - E_{down} = \hbar\omega$. So according to Boltzmann statistics we get for each stimulated transition the relation

$$\frac{n_{e,down}}{n_{e,up}} = \exp\left[\frac{E_{up} - E_{down}}{k_B T}\right] = \exp\left[\frac{\hbar\omega}{k_B T}\right] \qquad , \tag{6}$$

where $k_B$ is Boltzmann constant, and $T$ the absolute temperature. Summing over all transitions we get

$$\tilde{n}_{e,down} = \tilde{n}_{e,up} \exp\left(\frac{\hbar\omega}{k_B T}\right) \simeq \tilde{n}_{e,up}\left(1 + \frac{\hbar\omega}{k_B T}\right) \; ; \; \frac{\hbar\omega}{k_B T} \ll 1 \; ; \; \tilde{n}_{e,down} - \tilde{n}_{e,up} \simeq \tilde{n}_{e,up}\frac{\hbar\omega}{k_B T} \; . \quad (7)$$

Here $\tilde{n}_{e,down}$ and $\tilde{n}_{e,up}$ refer to the total number of electrons for stimulated emission and for stimulated absorption, respectively, and we assumed here the approximation that the photon energy is small relative to the thermal energy $k_B T$. We average the term $1/p$ over the Boltzmann distribution obtaining approximately

$$\langle (1/p) \rangle_{Boltz} \simeq \left(1/\sqrt{3k_B T m_e}\right) \simeq \left(\frac{1.628 \cdot 10^{26}}{\sqrt{T}}\right) \quad . \quad (8)$$

We multiply Eq. (4) by $\tilde{n}_{e,down} - \tilde{n}_{e,up} \simeq n_e \frac{\hbar\omega}{k_B T}$ and assume averaged value for $1/p$ according to Eq. (8). Then, we get for the average electromagnetic energy $U$ per unit volume, absorbed by the plasma per unit time

$$\left(\frac{dU}{dt}\right)_{ab} = C\left(\frac{1.628 \cdot 10^{26}}{\sqrt{T}}\right)\tilde{n}_{e,up}\frac{\hbar\omega}{k_B T} = \left[\frac{n_i Z^2}{\omega^3}\left(\frac{5.81 \cdot 10^{-10}}{T^{3/2}}\right)n_e\right]E^2 = \beta E^2 . \quad (9)$$

In Eq. (9) we used the approximation $\tilde{n}_{e,up} \simeq \tilde{n}_{e,down} \simeq n_e$. The proportionality constant $\beta$ gives the part of the electromagnetic energy represented by $E^2$ which is absorbed by the plasma per unit volume and unit time. Under the condition

$$\beta = \frac{n_i Z^2}{\omega^3}\left(\frac{5.81 \cdot 10^{-10}}{T^{3/2}}\right)n_e \ll 1 \quad , \quad (10)$$

the plasma becomes transparent. This condition depends on the parameters: $n_i, n_e, T$ and $\omega$ where for the present limiting case this condition is improved for lower ion and electron densities and higher temperatures. One should notice that the absorption of the plasma is proportional the multiplication $n_e n_i$ ($n_i \simeq n_e$) so that the absorption of the plasma decreases very much for smaller values of the electrons densities. The above calculations are made for slow electrons for which $Ze^2/4\pi\varepsilon_0 \hbar v > 1$. For Boltzmann statistics we use the approximation of average velocity $v = \sqrt{\frac{3k_B T}{m}}$ and then this condition can be written as

$$Z^2 \cdot 1.06 \cdot 10^5 > T\,^0K \quad . \quad (11)$$

So, for Carbon plasma ($Z = 6$) and Oxygen plasma ($Z = 8$) for $T = 10^5 \; ^0K$ it is a good approximation but for Hydrogen plasma ($Z = 1$) it is only a fair approximation.

## 3. The stability of stellar plasmas with low densities under Boltzmann statistics

We study here the stability properties of a star which includes plasma with relatively low densities. For simplicity of discussion we treat a star with spherical symmetry where its stability is related to gravitational forces without any other external perturbations e.g. magnetic fields etc. While this topic was studied under high electron densities using Fermi-Dirac statistics [Chandrasekhar 1967; Edington 1959; Kippenhan, Weigert &Weiss 2012] we study here the problem for low densities using Boltzmann statistics.

The gravitational force at a distance $r$ from the star center is due entirely to the mass $M_r$ interior to this distance:

$$g = GM_r / r^2 \quad , \tag{12}$$

where $G$ is the constant of gravitation. Assuming that $\phi$ is the gravitational potential then

$$g = -d\phi / dr \quad ; \quad \phi = -GM_r / r \quad . \tag{13}$$

According to the hydrostatic equation

$$dP = -g\rho dr \quad . \tag{14}$$

where $P$ is the star plasma pressure and $\rho$ the star density, both are functions of distance $r$ from the star center. Inserting Eq. (13) into Eq. (14) we get:

$$dP = \rho d\phi \quad . \tag{15}$$

This equation describes the increase of the plasma pressure as we move to larger values of $r$ balancing the attractive gravitational forces.

Assuming that the star plasma behaves as an ideal gas then the pressure $P$ is given by

$$P = n_e k_B T \quad . \tag{16}$$

Assuming also that the gradient of temperature is small relative to the gradient of the electron density $n_e$ (isothermal process) then we get:

$$dP = k_B T dn_e = \rho d\phi \quad . \tag{17}$$

If there are $\kappa$ nucleons for each electron, then the mass density is given approximately by

$$\rho = \kappa n_e m_N \quad . \tag{18}$$

Here $m_N \approx 1.67 \cdot 10^{-27} Kg$ is the mass of the nucleon and we assumed here $n_i = n_e$ where the number of ionized atoms $n_i$ is equal to the number $n_e$ of ionized electrons. Inserting Eq. (18) into Eq. (17):

$$\frac{dn_e}{d\phi} = \frac{\kappa m_N}{k_B T} n_e \to \ln(n_e) + C = \frac{\kappa m_N}{k_B T}\phi \quad , \quad C = -\ln(n_0) \to \ln\left(\frac{n_e}{n_0}\right) = \frac{\kappa m_N}{k_B T}\phi$$

$$n_e(r) = n_0 \exp\left[\left(\frac{\kappa m_N}{k_B T}\right)\phi(r)\right] \quad ; \quad \phi(r=0) = 0 \quad ; \quad n_e(r=0) = n_0 \tag{19}$$

This equation describes the decrease of the plasma density $n_e(r) = n_i(r)$ as a function of the change of the exponential function of the potential $\phi(r)$. Here $\phi(r)$ is changing from zero to large negative values, as function of the distance $r$ from the star center. Our aim in the following analysis is to find the change of $n_e(r) = n_i(r)$ as function of the distance $r$ from the star center. In these calculations we assume that $n_0$ is the density of ionized electrons in the star center which is taken as experimental parameter.

In order to find the dependence of the electron density $n_e(r)$ on the distance $r$ from the star center we need to take into account the Poisson equation for the potential $\phi$ which for a star with spherical symmetry has the form:

$$\frac{d^2\phi(r)}{dr^2} + \frac{2}{r}\frac{d\phi(r)}{dr} = -4\pi G \rho(r) \quad . \tag{20}$$

Here $\rho(r)$ is proportional to the ionized electron density $n_e(r)$, as given by Eq. (18), and $G$ is the gravitational constant. Substituting the relation $\rho = \kappa n_e m_N$ from Eq. (18) into Eq. (20) and using the relation (19) for $n_e(r)$ we get

$$\frac{d^2\phi(r)}{dr^2} + \frac{2}{r}\frac{d\phi(r)}{dr} = -4\pi G \kappa m_N n_e(r) = -4\pi G \kappa m_N n_0 \exp\left[\left(\frac{\kappa m_N}{k_B T}\right)\phi(r)\right] \quad . \tag{21}$$

We note that on the right side of Eq. (21) appears an exponential function of the potential $\phi(r)$ with a very small coefficient given by $\frac{\kappa m_N}{k_B T}$. One should notice also the coefficient before the exponential is a very small number for low electron densities. It is difficult to get explicit solutions to this equation, as series expansion of this exponential function converges very slowly for large values of $r$. We give in the Appendix an estimate to the main body of the electrons density profile by using series expansion for the exponential equation of Eq. (21). We prefer in the following analysis to transform Eq. (21) to differential equation for $n_e(r)$. It will give after some calculations the change of $n_e(r)$, from its initial value $n_e(r=0) = n_0$ at the star center (taken as experimental parameter), to smaller values as a function of the distance $r$ from the star center.

According to Eq. (19) we get:

$$\frac{\partial n_e(r)}{\partial r} = \left(\frac{\kappa m_N}{k_B T}\right)\frac{\partial \phi(r)}{\partial r} n_e(r) \rightarrow \frac{\partial \phi(r)}{\partial r} = \frac{1}{n_e(r)}\frac{\partial n_e(r)}{\partial r}\left(\frac{k_B T}{\kappa m_N}\right)$$

$$\frac{\partial^2 \phi(r)}{\partial r^2} = \left[-\frac{1}{n_e(r)^2}\left(\frac{\partial n_e(r)}{\partial r}\right)^2 + \frac{1}{n_e(r)}\frac{\partial^2 n_e(r)}{\partial r^2}\right]\left(\frac{k_B T}{\kappa m_N}\right) \tag{22}$$

Inserting Eq. (18) and the potential derivatives according to Eq. (22) into Eq. (20) we get:

$$\left[ -\frac{1}{n_e(r)^2}\left(\frac{\partial n_e(r)}{\partial r}\right)^2 + \frac{1}{n_e(r)}\frac{\partial^2 n_e(r)}{\partial r^2} + \frac{1}{n_e(r)}\frac{1}{r}\frac{\partial n_e(r)}{\partial r} \right]\left(\frac{k_B T}{\kappa m_N}\right) = -4\pi G \kappa m_N n_e \quad . \tag{23}$$

Multiplying Eq. (23) by $n_e(r)$ and rearranging the terms:

$$\frac{\partial^2 n_e(r)}{\partial r^2} + \frac{1}{r}\frac{\partial n_e(r)}{\partial r} - \frac{1}{n_e(r)}\left(\frac{\partial n_e(r)}{\partial r}\right)^2 = -4\pi G \frac{(\kappa m_N)^2}{k_B T} n_e(r)^2 \quad . \tag{24}$$

We define $\theta(r) = \dfrac{n_e(r)}{n_0}$, and divide Eq. (24) by $n_0$ then we get:

$$\frac{\partial^2 \theta(r)}{\partial r^2} + \frac{1}{r}\frac{(\partial \theta(r))}{\partial r} - \frac{1}{\theta(r)}\left(\frac{\partial \theta(r)}{\partial r}\right)^2 = -4\pi G \frac{(\kappa m_N)^2 n_0}{k_B T}\theta(r)^2 = -0.504 \cdot 10^{-46}\frac{n_0 \kappa^2 \theta(r)^2}{T} \quad ,$$

$$\theta(r) = \frac{n_e(r)}{n_0} \quad ; \quad n_0 = n_e(r=0) \tag{25}$$

We define

$$\xi = 4\pi G \frac{(\kappa m_N)^2 n_0}{k_B T} = 1.694 \cdot 10^{-40}\frac{n_0 \kappa^2}{T} \quad ; \quad x = r\sqrt{\xi} \simeq r \cdot 1.3 \cdot 10^{-20}\kappa\sqrt{\frac{n_0}{T}} \quad . \tag{26}$$

We divide Eq. (25), for $\theta(r)$ by $\xi$. Then, this equation can be written as:

$$\frac{\partial^2 \theta(x)}{\partial x^2} + \frac{1}{x}\frac{(\partial \theta(x))}{\partial x} - \frac{1}{\theta(x)^2}\left(\frac{\partial \theta(x)}{\partial x}\right)^2 = -\theta(x)^2 \quad . \tag{27}$$

Eq. (27) describes the change of the normalized electron density $n_e(x)/n_0 = \theta(x)$ as function of the normalized distance $x = r\sqrt{\xi}$ from the star center. This equation describes also the change of the mass density $\rho(Kg \cdot m^{-3}) = \kappa n_e m_N$ ($m_N = 1.67 \cdot 10^{-27} Kg$) as function of the distance from the star center. Here $m_N$ is the mass of nucleon and $\kappa$ is the ratio between the density of nucleons and that of electrons. (For completely ionized plasma: for Hydrogen plasma $\kappa = 2$, for Carbon $\kappa = 12$ and for Oxygen $\kappa = 16$ while for partial ionized plasma these numbers should increase inversely proportional to the part of ionization).

By neglecting the third term on the left side of Eq. (27) we obtain a differential equation which was solved numerically and given as the Lane-Emden equation with $n = 2$ [e.g. Mohan & Al-Bayaty 1980]. It shows the change of the value of $\theta(x) = 1$, for $x = 0$, at $r = 0$, to the value $\theta(x) = 0$ for $x = 1 = \sqrt{\xi} R$ where $R = 1/\sqrt{\xi}$ is approximately the star radius. . Such behavior occurs also for the solutions of the full Eq. (27), but since the third term, on the left side of Eq. (27), cannot be neglected the form of change of $\theta(x)$ is different from the Lane-Emden equation (with exponent $n = 2$). Like the Lane-Emden equation the range $x = 1$ describes approximately the range of values of $\theta(x)$, and substituting in Eq. (26) the value $x = 1 = R\sqrt{\xi}$ (where $\xi$ in the present conditions is extremely small number) we find that the radius $R$ of the star stretches over a very long distance which is of order $R = 1/\sqrt{\xi} \simeq \dfrac{7.7 \cdot 10^{19}}{\kappa}\sqrt{\dfrac{T}{n_0}}$ (m). Let us give more explicit estimations for the order of magnitudes of the star radius and star mass:

Order of magnitudes for the radius and mass of the low density star under Boltzmann statistics:

The radius of the star is then given approximately according to Eq. (26) by

$$R(star\ radius) \simeq \frac{1}{\sqrt{\xi}} \simeq \sqrt{\frac{k_B T}{4\pi G (\kappa m_N)^2 n_0}} \simeq \frac{7.7}{\kappa} \cdot 10^{19} \sqrt{\frac{T}{n_0}}\ (m) \qquad . \qquad (28)$$

We get here the interesting result that the radius of the star $R$ increases inversely proportional to the square root of its electron density in the star center.

The total mass $M_{star}$ of the star is proportional to its volume, and to the average mass density given according to Eq. (18) as $\kappa m_N \langle n_e \rangle$ where $\langle n_e \rangle$ is the average electron density given approximately by $\langle n_e \rangle \approx \dfrac{n_0}{2}$. Then we get:

$$M_{star}(star\ mass) \simeq \frac{4\pi}{3} R^3 \kappa m_N \frac{n_0}{2} = \frac{2\pi}{3 m_N^2} \left(\frac{k_B}{4\pi G}\right)^{3/2} \frac{T^{3/2}}{\kappa^2 \sqrt{n_0}} = 5.76 \cdot 10^{32} \frac{T^{3/2}}{\kappa^2 \sqrt{n_0}}\ (Kg) \quad . \quad (29)$$

Here we substituted the value of the star radius according to Eq. (28) and inserted an averaged value for the electron density $\langle n_e \rangle$ which is proportional to its density $n_o$ in the star center. We find that the star mass is increasing with temperature as it is proportional to $T^{3/2}$. We find also the astonishing result that the star mass is inversely proportional to the square root of the density of the electrons in its center due to increase in the volume proportional to $R^3$.

The above effects can be summarized as follows: for low densities plasma the balance between the ideal gas radiation pressure and gravitational forces leads to extremely large value for the star radius so that although the mass density is quite small the star volume is extremely large so that it leads to very large star mass with strong gravitational forces. We demonstrate these effects in the next section by numerical calculations for three examples in which: a) $n_0 = 10^{26}\ (m^{-3})$, b) $n_0 = 10^{23}\ (m^{-3})$, c) $n_0 = 10^{20}\ (m^{-3})$ and for all these cases with the temperature $10^5,\ ^0K$. These results are decreasing somewhat for lower temperatures as the star radius is proportional to $\sqrt{T}$ and the star mass is proportional to $T^{3/2}$. We demonstrate our results by explicit calculations only for 3 examples. The present effects become even much stronger for lower electron densities.

## 4. Numerical Calculations

We demonstrate our results in the following 3 examples.

**Star with electron density** $n_0 = 10^{26} \left( m^{-3} \right)$ **, temperature** $10^5, {}^0K$

According to Eq. (28) we get:

$$R(star\ radius) \simeq \frac{7.7}{\kappa} \cdot 10^{19} \sqrt{\frac{10^5}{10^{26}}}\ (m) = \frac{2.435 \cdot 10^9}{\kappa}(m) = \frac{2.435 \cdot 10^6}{\kappa}(km) \quad . \tag{30}$$

For completely ionized Hydrogen plasma $\kappa = 2$, so that the radius of such star is given approximately by: $1.22 \cdot 10^6\ (km)$. It has a radius which is larger approximately by a factor 1.75 relative to solar radius. The radius of the present star becomes smaller for Carbon or Oxygen plasma for which $\kappa = 12$ or 16, respectively. The radius is also decreased for lower temperature as according to Eq. (28) it is proportional to $\sqrt{T}$. We have to take into account that for partial ionization the parameter $\kappa$ increases inversely proportional to the part of ionization and thus the star radius is decreased further.

Substituting in (29) $T = 10^5, \ n_0 = 10^{26},$ then we get:

$$M_{star}(star\ mass) \simeq 5.76 \cdot 10^{32} \frac{10^{15/2}}{\kappa^2 \sqrt{10^{26}}}(Kg) = \frac{1.82 \cdot 10^{27}}{\kappa^2}(Kg) \quad . \tag{31}$$

This mass, for completely ionized Hydrogen plasma ($\kappa = 2$) is smaller, approximately by factor 4500 relative to the solar mass. For smaller temperature this mass decreases as it is proportional to $T^{3/2}$. It also decreases for larger values of $\kappa$ for Carbon and Oxygen plasma and also for partial ionization. The above effects become much stronger for smaller densities as analyzed in the following examples.

For completely ionized plasma we assume in Eq. (10) the approximation $n_e = n_i \simeq n_0$ (neglecting neutral atoms) and for $n_0 = 10^{26}; \ T = 10^5\ {}^0K$, the transparency condition for the plasma becomes:

$$\beta \simeq \frac{1.83 \cdot Z^2 \cdot 10^{35}}{\omega^3} \ll 1 \quad . \tag{32}$$

For hydrogen plasma which is the most common plasma we have $Z = 1$, and the condition for transparency is given by $\omega^3 \gg 1.83 \cdot 10^{35}$. So, for the optical region and near infrared the Hydrogen plasma is transparent but for the far infrared the absorbance becomes large. Also the absorbance becomes much larger for plasmas for which $Z^2 \gg 1$. The condition for transparency is improved very much for lower densities, as treated by the following examples.

**Star with electron density** $n_0 = 10^{23} \left( m^{-3} \right)$ **, temperature** $10^5, {}^0K$

Substituting in Eq. (28) $n_0 = 10^{23}$, $T = 10^5$ we get:

$$R(\text{star radius}) \simeq \frac{7.7}{\kappa} \cdot 10^{19} \sqrt{\frac{10^5}{10^{23}}} \ (m) \simeq \frac{7.7 \cdot 10^{10}}{\kappa} (m) = \frac{7.7 \cdot 10^7}{\kappa} (km) \quad . \tag{33}$$

For completely ionized Hydrogen plasma $\kappa = 2$, and such star has a radius which is larger approximately by a factor 55, relative to the solar radius ! This radius becomes smaller for plasmas for which $\kappa > 2$ and for lower temperatures.

Substituting in Eq. (29) the values $T = 10^5$, $n_0 = 10^{23}$, then we get:

$$M_{star}(\text{star mass}) \simeq \frac{5.7 \cdot 10^{32} 10^{15/2}}{\kappa^2 \sqrt{10^{23}}} \simeq \frac{5.7 \cdot 10^{28}}{\kappa^2} \ (Kg) \quad . \tag{34}$$

So, for completely ionized hydrogen plasma for which: $\kappa = 2$, the present star mass is smaller approximately by factor 143 than the solar mass. For plasmas with values of $\kappa$ larger than 2 this value can be reduced but in any case it gives very large mass for low electron density star due to the extremely large volume.

For completely ionized plasma we assume in Eq. (10) the approximation $n_e = n_i \simeq n_0/2$ (neglecting neutral atoms) and for $n_0 = 10^{23}$; $T = 10^5$, $^0K$, the transparency condition for the plasma becomes

$$\beta \simeq 1.83 \frac{Z^2 \cdot 10^{29}}{\omega^3} \ll 1 \quad . \tag{35}$$

For hydrogen plasma we have $Z = 1$, and the plasma becomes completely transparent for $\omega > 10^{12} (\sec^{-1})$ which includes the entire significant electromagnetic spectrum. The absorbance becomes larger for plasmas for which $Z^2 \gg 1$.

**Star with electron density $n_0 = 10^{20} \ (m^{-3})$, temperature $10^5$, $^0K$**

Substituting in Eq. (28) $n_0 = 10^{20}$, $T = 10^5$ we get:

$$R(\text{star radius}) \simeq \frac{7.7}{\kappa} \cdot 10^{19} \sqrt{\frac{10^5}{10^{20}}} \ (m) \simeq \frac{2.435 \cdot 10^{12}}{\kappa} (m) = \frac{2.435 \cdot 10^9}{\kappa} (km) \quad . \tag{36}$$

For completely ionized Hydrogen plasma: $\kappa = 2$, so that the radius of such star has a radius which is larger approximately by a factor 1750 relative to the Solar Radius !. This radius becomes smaller for plasmas for which $\kappa > 2$ and for lower temperatures.

Substituting in Eq. (29) the values $T = 10^5$, $n_0 = 10^{20}$, then we get:

$$M_{star}(\text{star mass}) \simeq \frac{5.76 \cdot 10^{32} 10^{15/2}}{\kappa^2 \sqrt{10^{20}}} \simeq \frac{1.82 \cdot 10^{30}}{\kappa^2} \quad . \tag{37}$$

So, for completely ionized hydrogen plasma for which: $\kappa = 2$, the present star mass is smaller approximately by factor 4.5 than the solar mass !. For plasmas with values of $\kappa$ larger than 2 this value can be reduced, but for electrons densities which are smaller than that of $n_0 = 10^{20}$ the star mass becomes larger. In any case it gives large mass for low electron density star due to the extremely large volumes. For completely ionized plasma we assume in Eq. (10) the approximation $n_e = n_i \simeq n_0 / 2$ (neglecting neutral atoms) and for $n_0 = 10^{20}$; $T = 10^5\ ^0K$, the transparency condition for the plasma becomes

$$\beta \simeq 1.83 \frac{Z^2 \cdot 10^{23}}{\omega^3} \ll 1 \quad . \tag{38}$$

For this case Hydrogen, Carbon and Oxygen plasmas become transparent over all the entire significant electromagnetic spectrum.

We demonstrated the present ideas by analyzing 3 examples. The present effects become even stronger, if the density of electrons is reduced further. We assumed in our analysis a star with spherical symmetry under the opposing forces: gravitational forces and radiation pressure for plasma behaving as an ideal gas. The assumption of a star with spherical symmetry is assumed here mainly due to mathematical convenience. I expect that similar effects will occur in low density star plasmas in which the plasmas will have different various structures. It seems that the present analysis is relevant to dark matter.

## 5. Summary Discussions and Conclusions

In the present work we have shown that the absorption of low ionized stellar plasmas with one ionic component with $Z$ atomic number, satisfying Boltzmann statistics, in which the condition $Ze^2 / 4\pi\varepsilon_0 \hbar v > 1$ is satisfied, can be treated approximately by theories about stimulated Bremsstrahlung. The stimulated emission minus stimulated absorption leads to inverse stimulated absorption which can be neglected under the condition given by Eq. (10). As this condition is proportional to multiplication of the density of electrons $n_e$ times the density of ionized ions $n_i$ assuming $n_i \simeq n_e$ (especially correct for Hydrogen ionized plasma) the plasma becomes transparent for small such densities. This condition is improved for smaller values of the atomic number $Z$ (Z=1 for Hydrogen) and for higher temperatures. The mass density $\rho$ of a star based on the present conditions is given by Eq. (18) as $\rho = \kappa n_e m_N$ where $m_N$ is the mass of nucleon and $\kappa$ is the number of nucleons per one electron (for Hydrogen completely ionized plasma $\kappa = 2$). We find that for the present transparent plasmas the mass density is very small but, as analyzed in the present article, the stability conditions for such plasmas leads to extremely large volumes so that gravitational forces become very large. In the present work we assumed that the star plasma behaves as an ideal gas where the pressure $P = n_e k_B T$ was given by Eq. (16), where $k_B$ is the Boltzmann constant. The stability of the star with low electron densities is obtained by considering it to be in equilibrium between two opposing forces: The repulsive ideal gas electron pressure and the attractive gravitational force. For a star with spherical symmetry the balance between these opposite forces leads to Eq. (19). In this equation the density of electrons is given by an exponential function of the potential with a small coefficient in the exponent. In principle, this equation can be solved by taking into account the Poisson equation for the potential given in Eq. (20). Since these equations are difficult to solve we preferred in the present analysis to transform

Eq. (20) to a differential equation for the density of electrons $n_e(r)$ as function of the distance $r$ from the star center. After some calculations we obtained the differential equation for $n_e(r)$ in Eq. (23). After some additional transformations we obtained the differential equation (27) for the normalized electron density $\theta(r) = n_e(r)/n_0$ as function of the normalized distance $x = r\sqrt{\xi}$ of Eq. (26). We estimated the star radius R as $1/\sqrt{\xi}$ where $\xi$ is a very small radius given by Eq. (26). Using this estimate we made numerical calculations for the star radius and mass in section 4, for 3 examples for which the electron density in the mass center is given by $a)\ n_0 = 10^{26}\ (m^{-3}),\ b)\ n_0 = 10^{23}\ (m^{-3}),\ c)n_0 = 10^{20}\ (m^{-3})$. We demonstrate by these calculations the interesting result by which the radius $R$ of the star increases inversely proportional to the square root of the electron density $n_0$ at the star center as given by Eq. (28). The mass density of the star increases proportional to the electron density as given by Eq. (18). The star volume, however, increases proportional to $R^3$ leading the total mass of the star to increase inversely proportion to the square root of the electron density $n_0$ at the star center as given by Eq. (29)! We added in the Appendix a calculation for the star radius by using a series expansion of the exponential of Eq. (19). This calculation gives an order of magnitude for the star radius which is similar to that given by Eq. (28).

Usually one would think that bremsstrahlung phenomena cannot be ignored when electrons are colliding with ions but we should take into account that, under the conditions presented in the present work, bremsstrahlung stimulated emission involves also bremsstrahlung stimulated absorption so that the summation of these two opposing effects leads to inverse stimulated absorption as given by Eq. (9). Since this absorption is proportional to the multiplication of the density of electrons $n_e(r)$ by the ion densities $n_i(r)$ (where $n_e(r) \simeq n_i(r)$) this absorption is reduced very much for lower densities and vanishes over all the entire significant electromagnetic spectrum, for $n_e(r) \leq 10^{20}\ (m^{-3})$, as demonstrated by the numerical calculations in section 4. Trying to relate the present work to certain phenomena about dark matter we need to use the following considerations. In the above analysis we obtained extremely large volumes for plasma with low electron density, where we got the result that as we decrease the density of the electrons the volumes increases very much. The astonishing result from the present analysis is that a star with low mass density extends over extremely large volumes so that the total mass is very large and gravitational forces become very significant. These phenomena which seem to be quite strange follow from the fact that in order that the gravitational forces of low density star will be in equilibrium with the radiation pressure of ideal electron gas they should extend on larger and larger volumes as the mass density is decreased. The present analysis of a star with spherical phenomena was made for simplicity of calculations but it does not imply that such plasmas will have spherical symmetry since similar effects will occur on other different structures of the plasmas in which the explicit calculation should be different. Finally, following an article [Ben-Aryeh 1967] which was published quite long ago, the emissivity (the ratio between the experimental radiation and the black body radiation) is equal to the absorbance so that for transparent plasma the emissivity vanishes and we will not observe any radiation emitted from such plasma.

## APPENDIX

We give here an alternative derivation for the star radius starting from Eq. (21). We substitute, in Eq. (21) the following definitions:

$$y^2 = 4\pi G \kappa m_N n_0 r^2 \quad ; \quad z = \frac{\kappa m_N}{k_B T} \tag{A1}$$

and express the potential function as a series expansion. Then this equation can be written as:

$$\frac{d^2\phi(y)}{dy^2} + \frac{2}{y}\frac{d\phi}{dy} = -\left\{1 + z\phi(y) + \frac{z^2\phi(y)^2}{2} + \frac{z^3\phi(y)^3}{3!} + \cdots\right\}. \tag{A2}$$

The solution for $\phi(y)$ can be estimated by using the following series expansion:

$$\phi(y) = 0 - ay - by^2 - cy^3 - dy^4 \cdots . \tag{A3}$$

Inserting Eq. (A3) into Eq. (A2) and performing the derivatives of $\phi(y)$ we get:

$$\left(-\frac{2a}{y} - 6b - 12cy - 32dy^2 + \cdots\right) + \left\{1 + \sum_{n=1,2,3,\cdots} \frac{\left[z(ay + by^2 + cy^3 + \cdots)\right]^n}{n!}\right\} = 0. \tag{A4}$$

We obtain the coefficients $a, b, c, d, \cdots$ corresponding respectively to $n = -1, 0, 1, 2, 3$ by equating to zero the sum of terms with the same exponent of $y$. We get:

$$a = c = e = 0 \quad ; \quad b = 1/6 \quad ; \quad d = -\frac{1}{120}z \quad , \quad \cdots . \tag{A5}$$

Here we used the comparisons for $y^n$ only for $n = -1, 0, 1, 2, \cdots$ but the comparison can be made for larger values of $n$ with somewhat lengthy calculations.

Since $z$ is a small number we get for the solution of $\phi(y)$:

$$\phi(y) \simeq -(1/6)y^2 = -\frac{2}{3}\pi G \kappa m_N n_0 r^2 = \phi(r) . \tag{A6}$$

Substituting $\phi(y)$ of Eq. (A6) into Eq. (19) for $n_e(r)$ we get:

$$n_e(r) = n_0 \exp\left[-\frac{2}{3}\frac{\pi G (\kappa m_N)^2 n_0 r^2}{k_B T}\right] . \tag{A7}$$

We get that under the condition $r^2 = \frac{3}{2}\frac{k_B T}{\pi G (\kappa m_N)^2 n_0} = R^2$ the density of the star $n_e(r)$ is reduced to $\frac{1}{e}n_0$ so that the radius $R$ of this star can be estimated as

$$R = \sqrt{\frac{3}{2} \frac{k_B T}{\pi G (\kappa m_N)^2 n_0}} \qquad . \tag{A8}$$

So this estimation for the star radius is larger by factor $\sqrt{6} \simeq 2.45$ relative to the estimation made in Eq. (28) but both estimations lead to the same orders of magnitudes.